\documentclass[fleqn,10pt]{wlscirep}
\usepackage[utf8]{inputenc}
\usepackage[T1]{fontenc}
\usepackage{chemformula}
\let\ce\ch
\usepackage{mathtools}
\usepackage{physics}
\usepackage{floatrow}
\usepackage{subcaption}
\usepackage{hyperref}

\title{Entangled radicals may explain lithium effects on hyperactivity}

\author[1,2,3,4*]{Hadi Zadeh-Haghighi}
\author[1,2,3,4*]{Christoph Simon}

\affil[1]{Department of Physics and Astronomy, University of Calgary, Calgary, AB, T2N 1N4, Canada}
\affil[2]{Institute for Quantum Science and Technology, University of Calgary, Calgary, AB, T2N 1N4, Canada}
\affil[3]{Quantum Alberta, and Hotchkiss Brain Institute, University of Calgary, Calgary, AB, T2N 1N4, Canada}
\affil[4]{Hotchkiss Brain Institute, University of Calgary, Calgary, AB, T2N 1N4, Canada}

\affil[*]{hadi.zadehhaghighi@ucalgary.ca, csimo@ucalgary.ca}

\begin{abstract}

It is known that bipolar disorder and its lithium treatment involve the modulation of oxidative stress. Moreover, it has been observed that lithium's effects are isotope-dependent. Based on these findings, here we propose that lithium exerts its effects by influencing the recombination dynamics of a naturally occurring radical pair involving oxygen. We develop a simple model inspired by the radical-pair mechanism in cryptochrome in the context of avian magnetoreception and xenon-induced anesthesia. Our model reproduces the observed isotopic dependence in the lithium treatment of hyperactivity in rats. It predicts a magnetic-field dependence of the effectiveness of lithium, which provides one potential experimental test of our hypothesis. Our findings show that Nature might harness quantum entanglement for the brain's cognitive processes.

\end{abstract}
\begin{document}

\flushbottom
\maketitle
\thispagestyle{empty}

\section*{Introduction}
The human brain is a magnificent system with highly complex functionalities such as learning, memory, emotion, and subjective experience, each of which is \textit{mood} dependent \cite{MartnezArn2004}. Everyday millions of patients all over the world take psychiatric pharmaceuticals to stabilize their mood, yet the underlying mechanisms behind these medications remain largely unknown\cite{Grande2016}.\par 

Bipolar disorder (BD) is a devastating mental illness which affects 2-4\% of the world population \cite{Geddes2013}. As its name implies, it entails two distinct oscillating and opposing states, a state of energy and hyperactivity (the manic phase), and a state of low energy (depression) \cite{McIntyre2020}. Lithium (Li) administration is the first-line treatment of bipolar illness \cite{Geddes2013,Malhi2017,MachadoVieira2009,Khairova2011}. Despite the common clinical use of this drug, the mechanism by which it exerts its effects remains elusive \cite{Roux2017,Harwood2004}. Here, based on experimental findings, we propose a mechanism for Li's therapeutic effects on BD.\par

Studies in the 1980s showed that administering Li results in different parenting behaviors and potentially delayed offspring development in rats \cite{Sechzer1986}. Although these findings weren't quantitative, it was reported that different Li isotopes have different impacts. Recently, a new study was conducted by Ettenberg \textit{et al.} \cite{Ettenberg2020} demonstrating an isotope effect of Li on the manic phase in rats. Li has two stable isotopes, \ce{^{6}Li} and \ce{^{7}Li}, which carry different nuclear spin angular momentum, $I_6=1$ and $I_7=3/2$, respectively. In that study, sub-anesthetic doses of ketamine were administered to induce hyperactivity which was then treated with lithium. The findings of that work indicate that \ce{^{6}Li} produces a longer suppression of hyperactivity in an animal model of mania compared to \ce{^{7}Li}. \par 

Moreover, there is accumulating evidence indicating that BD \cite{Andreazza2008,Yumru2009,Salim2014,MachadoVieira2007,Ng2008,Lee2013,Brown2014,Berk2011} and its Li treatment \cite{Khairova2011,frey2006effects,MachadoVieira2007,deSousa2014} are associated with oxidative stress, which is an imbalance between production and accumulation of radical oxygen species (ROS) in cells and tissues and the ability of a biological system to detoxify these reactive products, essential for governing life processes \cite{Sies2017}. It therefore seems  pertinent to explore the relationship between nuclear spin effects and oxidative stress in lithium mania therapy.\par

Any proposed mechanism for Li's BD treatment should embrace these two facts: Li's effect on BD exhibits isotopic dependence, and BD is connected to abnormalities in oxidative stress. In other words, the balance of naturally occurring free radicals is disrupted in BD patients, and lithium's different isotopes appear to alter the free radical formations differently. It is therefore possible that nuclear spin properties might be the key for the differential effects of the two lithium isotopes in BD. Here, we propose that lithium affects naturally occurring radical pairs (RPs) in a way that is modulated by the hyperfine interaction (HFI). \par

Spins can play crucial roles in chemical reactions even though the energies involved are orders of magnitude smaller than the thermal energy, $k_BT$ \cite{Steiner1989,Hayashi2004}. It has been known since the 1970s that external magnetic fields and nuclear spins can alter the rates and product yields of certain chemical reactions \cite{Schulten1976,Brocklehurst1974}. The key ingredients are RPs created simultaneously, such that the two electron spins, one on each radical, are entangled. Organic radicals are typically created in singlet (S) or triplet (T) entangled states by a reaction that conserves electron spin. Any spin in the vicinity or an external magnetic field can alter the extent and timing of the S-T interchange and hence the yields of products formed spin-selectively from the S and T states \cite{Jones2016,Timmel1998}. \par

Over the past decades, it has been proposed that that quantum physics could help answer unsolved questions in life science \cite{Ball2011,mcfadden2016life,Kim2021}. The above-described Radical Pair Mechanism (RPM) is one of the most well-established models in quantum biology; it could give a promising explanation for how migratory birds can navigate by means of Earth’s magnetic field \cite{Hore2016,Mouritsen2018}. The application of RPM has begun to gain momentum in numerous fields of research \cite{Hore2020}. Most recently, Smith \textit{et al.}, inspired by the the RPM explanation of avian magnetoreception, have shown that the RPM could play a role in xenon-induced anesthesia, which exhibits isotopic dependence \cite{Smith2021}.\par

In the context of avian magnetoreception, the RPM is thought to involve the protein cryptochrome (Cry) \cite{Ritz2000}, which contains the flavin adenine dinucleotide (FAD) cofactor. It is known that in Cry RPs can be in the form of flavosemiquinone radical (\ch{FAD^{.-}}) and terminal tryptophan radical (\ch{TrpH^{.+}}) \cite{Giovani2003,Hore2016,Hong2020}. However, considerable evidence suggests that the superoxide radical, \ch{O2^{.-}}, can be an alternative partner for the flavin radical, such that \ch{FADH^{.}} and \ch{O2^{.-}} act as the donor and the acceptor, respectively \cite{Mller2011,Romero2018,Chaiyen2012,Mondal2019}. This is motivated by the fact that the magnetically sensitive step in the reaction scheme occurs in the dark \cite{Wiltschko2016}, with required prior exposure to light, and that a strongly asymmetric distribution of hyperfine couplings across the two radicals results in  a stronger magnetic field effect than a more even distribution \cite{Hogben2009}. Furthermore, Zhang \textit{et al.} \cite{Zhang2021} recently showed that a magnetic field much weaker than that of the Earth attenuates adult hippocampal neurogenesis and cognition and that this effect is mediated by ROS. Additionally, it has been shown that the biological production of ROS \textit{in vivo} can be influenced by oscillating magnetic fields at Zeeman resonance, indicating coherent S-T mixing in the ROS formation \cite{Usselman2016}. \par 

BD is characterized by shifts in energy, activity, and mood and is correlated with disruptions in circadian rhythms \cite{Abreu2015,Takahashi2008} and abnormalities in oxidative stress \cite{Yumru2009,Salim2014,MachadoVieira2007,Ng2008,Andreazza2008,Lee2013,Brown2014,Berk2011}. It is known that Li influences the circadian clock in humans, and circadian rhythms are disrupted in patients with BD for which Li is a common treatment \cite{Yin2006,Li2012}. Yet, the exact mechanisms and pathways behind this therapy are under debate. However, it has been shown that Li acts directly on the mammalian suprachiasmatic nucleus (SCN), a circadian pacemaker in the brain \cite{abe2000lithium,Osland2010}. Cry's are necessary for SCN's development of intercellular networks that subserve coherent rhythm expression\cite{Welsh2010}. Further, it has also been reported that Cry2 is associated with BD \cite{Lavebratt2010}. In addition, Cry's essential role has been discovered in axon outgrowth in low intensity repetitive transcranial magnetic stimulation (rTMS) \cite{Dufor2019}. It thus seems that Cry might also have vital roles in the brain's functionalities related to BD. The [\ch{FADH^{.}}... \ch{O2^{.-}}] RP formation in Cry is consistent with evidence that Li treatment \cite{Khairova2011,frey2006effects,MachadoVieira2007,deSousa2014} and BD \cite{Yumru2009,Salim2014,MachadoVieira2007,Ng2008,Andreazza2008,Lee2013,Brown2014,Berk2011} are associated with oxidative stress implying a key role for radicals in BD and Li effects. Motivated by this reasoning, in the present work, we specifically focus on the Cry pathway for Li therapy, however, this is not the only way in which radical pairs could play a role in lithium effects on BD. Alternative paths will be discussed briefly in the Discussion section. \par

\begin{figure}[ht!]
    \centering
    \includegraphics[width=0.4\textwidth]{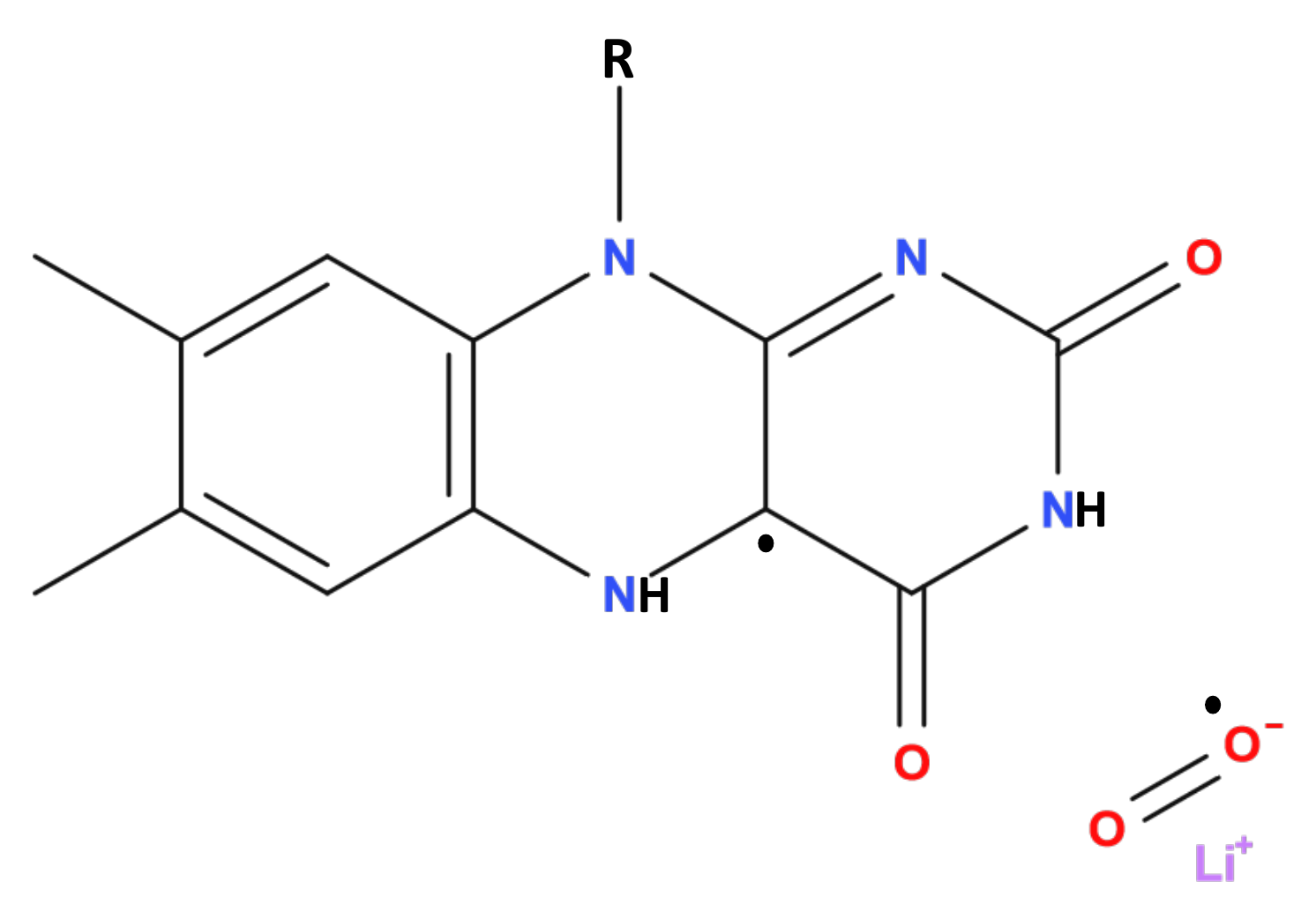}
   \caption{Flavinsemiquinone [\ch{FADH^{.}}] and lithium superoxide  [\ch{Li^{+}}-\ch{O2^{.-}}] radical pair. The radical pair undergoes intersystem crossing between singlet and triplet entangled states.}
    \label{fig:schem}
\end{figure}

Here, we propose that the RPM could be the underlying mechanism behind the isotope effects of Li treatment for hyperactivity. We propose that Li’s nuclear spin modulates the recombination dynamics of S-T interconversion in the naturally occurring RPs in the [\ch{FADH^{.}}... \ch{O2^{.-}}] complex, and due to the distinct nuclear spins, each isotope of Li influences these dynamics differently, which results in different therapeutic effects, see Fig. \ref{fig:schem}.\par

Let us note that an alternative interpretation has also been proposed for the isotope dependence in lithium's effect on mania. Fisher has suggested that phosphorus nuclear spins could be entangled in networks of Posner molecules, ${Ca}_{9}({PO}_4)_6$, which could form the basis of a quantum mechanism for neural processing in the brain \cite{Fisher2015}. Replacement of calcium at the center of a Posner molecule with another cation can further stabilize the Posner molecule \cite{Swift2018}, where each isotope of the cation, due to their different nuclear spins, has a different impact. 
This also provides a potential interpretation for the lithium isotope dependence treatment for mania. However, this model demands more supporting evidence and has been recently challenged experimentally \cite{Chen2020}. Moreover, our proposed explanation based on the RPM allows us to make connections with the essential roles of ROS \cite{Zhang2021} and Cry \cite{Dufor2019} in the brain's neuronal activity and in the biological pathways for lithium effects on BD. In the following, we aim to probe the possibility of RPM  for lithium treatment and analyze the parameter values that are required in order to explain the isotopic effect of Li's BD treatment observed by Ettenberg \textit{et al.} \cite{Ettenberg2020}.\par

\section*{Results}

\subsection*{Lithium Effectiveness on hyperactivity and RPM}
\subsubsection*{Quantifying effectiveness}
Ettenberg \textit{et al.} \cite{Ettenberg2020} recorded the effects of lithium treatments on ketamine-induced locomotor activity of rats. They conducted the test over 60 min sessions beginning immediately after ketamine injection. After administering of \ce{^{6}Li} and \ce{^{7}Li}, they measured the mean traveled distance with the standard error of the mean (SEM) for every five minutes for a group of 16 rats for each isotope treatment. They observed that \ce{^{6}Li} treatment exhibited significantly greater and more prolonged reductions in hyperactivity compared to \ce{^{7}Li}. Here we define the cumulative traveled distance in 60 minutes $TD_6$ and $TD_7$, respectively, for \ce{^{6}Li} and \ce{^{7}Li}. The total traveled distance ratio $TD_r$ is defined such that $TD_r = TD_7/TD_6$, see Table \ref{tab:Li-exp}. We derived the uncertainties based on the reported mean ($\pm$SEM) hyperlocomotion from Ettenberg \textit{et al.} \cite{Ettenberg2020} using standard error propagation.

\begin{table}[tbhp]
\centering
\caption{\label{tab:Li-exp}Li isotopic nuclear spins and total traveled distance in 60 minutes for \ce{^{6}Li} and \ce{^{7}Li}, $TD_6$ and $TD_7$, respectively, taken form the work of Ettenberg \textit{et al.} \cite{Ettenberg2020}. $TD_r$ is the total traveled distance ratio.} 
\begin{tabular}{lcc}
Treatment &  Total Traveled Distance (cm)& Uncertainty\\
\hline
\ce{^{6}Li} ($I=1$) & $TD_6=20112$ & $\pm 4.62\%$ \\
\ce{^{7}Li} ($I=3/2$) & $TD_7=25630$ & $\pm 3.83\%$ \\
\hline
  &$TD_r=TD_7/TD_6=1.3\pm6\%$ &
\end{tabular}
\end{table}

\subsubsection*{RPM model}

The RP model developed here is used to reproduce the effectiveness of Li for treatment of hyperactivity based on alteration in the triplet yield for different isotopes.  Here, we propose that Li may interact with the RP system of FADH and superoxide. The correlated spins of RP are assumed to be in [\ch{FADH^{.}}... \ch{O2^{.-}}], where the unpaired electron on \ch{O2^{.-}} couples with the Li nuclear spin, see Fig. \ref{fig:schem}.\par

In this model, we consider a simplified form of interactions for the RPM by only including Zeeman and HF interactions \cite{Efimova2008,Hore2016}. Given the likely randomized orientation of the relevant proteins, for the HFIs, we only consider the isotropic Fermi contact contributions. In our calculations, we assume that the unpaired electron on \ch{FADH^{.}} couples only with the isoalloxazine nitrogen nucleus, which  has the largest isotropic HF coupling constant (HFCC) among all the atoms in FAD  \cite{Maeda2012}, following the work of Hore \cite{Hore2019}, and that the other unpaired electron on [\ch{Li^{+}}-\ch{O2^{.-}}] couples with the lithium nucleus. The Hamiltonian for our RP system reads as follows:    

\begin{equation}
    \hat{H}=\omega \hat{S}_{A_{z}}+a_1 \mathbf{\hat{S}}_A.\mathbf{\hat{I}}_1+\omega \hat{S}_{B_{z}}+a_2 \mathbf{\hat{S}}_B.\mathbf{\hat{I}}_2,
\end{equation}
where $\mathbf{\hat{S}}_A$ and $\mathbf{\hat{S}}_B$ are the spin operators of radical electron A and B, respectively, $\mathbf{\hat{I}}_1$ is the nuclear spin operator of the isoalloxazine nitrogen of  \ch{FADH^{.}}, $\mathbf{\hat{I}}_2$ is the nuclear spin operator of the Li nucleus, $a_1$ is the HFCC between the isoalloxazine nitrogen of  \ch{FADH^{.}} and the radical electron A ($a_1=523.3 \mu T$ \cite{Maeda2012}), $a_2$ is the HFCC between the Li nucleus and the radical electron B, and $\omega$ is the Larmor precession frequency of the electrons due to the Zeeman effect.

\subsubsection*{DFT}
\begin{figure}[ht!]
    \centering
    \includegraphics[width=0.275\textwidth]{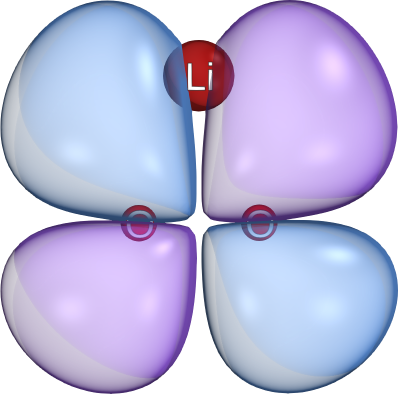}
   \caption{The highest occupied molecular orbital of [\ch{Li^{+}}-\ch{O2^{.-}}]. Imaged rendered using IboView v20150427 2
(http://www.iboview.org).}
    \label{fig:homo}
\end{figure}

We use density functional theory (DFT) to determine the reasonable range for the Li HFCC. Our DFT calculations show that the unpaired electron of [\ch{Li^{+}}-\ch{O2^{.-}}] is bound. The the highest occupied molecular orbital (HOMO) is shown in Fig. \ref{fig:homo}. The resulting Mulliken charge and spin population of the [\ch{Li^{+}}-\ch{O2^{.-}}] complex indicates that the unpaired electron resides primarily on the \ch{O2} molecule but is extended slightly onto the lithium atom, see Table \ref{tab:mulliken}. \par

\begin{table}[tbhp]
\centering
\caption{\label{tab:mulliken}Mulliken charge and spin population of [\ch{Li^{+}}... \ch{O2^{.-}}].} 
\begin{tabular}{ccc}
Atom & Charge Population &Spin Population\\
\hline
O & -0.244409 & 0.524734 \\

O & -0.244398 &   0.524744 \\

Li & 0.488807  & -0.049479 \\
\hline
Sum & 	0 & 1 \\
\end{tabular}
\end{table}

Using different DFT functionals and basis-sets, we find a range of values for the HFCC of Li nucleus at the distance of $\sim 1.6 \text{\AA}$ from \ch{O2^{.-}}, which is close to the values from other studies \cite{Lau2011}, $a_2 \in [157.7, 282]  \mu$T. The HFCC values between the unpaired electron on \ch{FADH^{.}} and the isoalloxazine nitrogen nuclear spin are taken from Maeda \textit{et al.} \cite{Maeda2012}. \par

\begin{figure}%
\centering
\begin{subfigure}{.7\textwidth}
\includegraphics[width=0.7\textwidth]{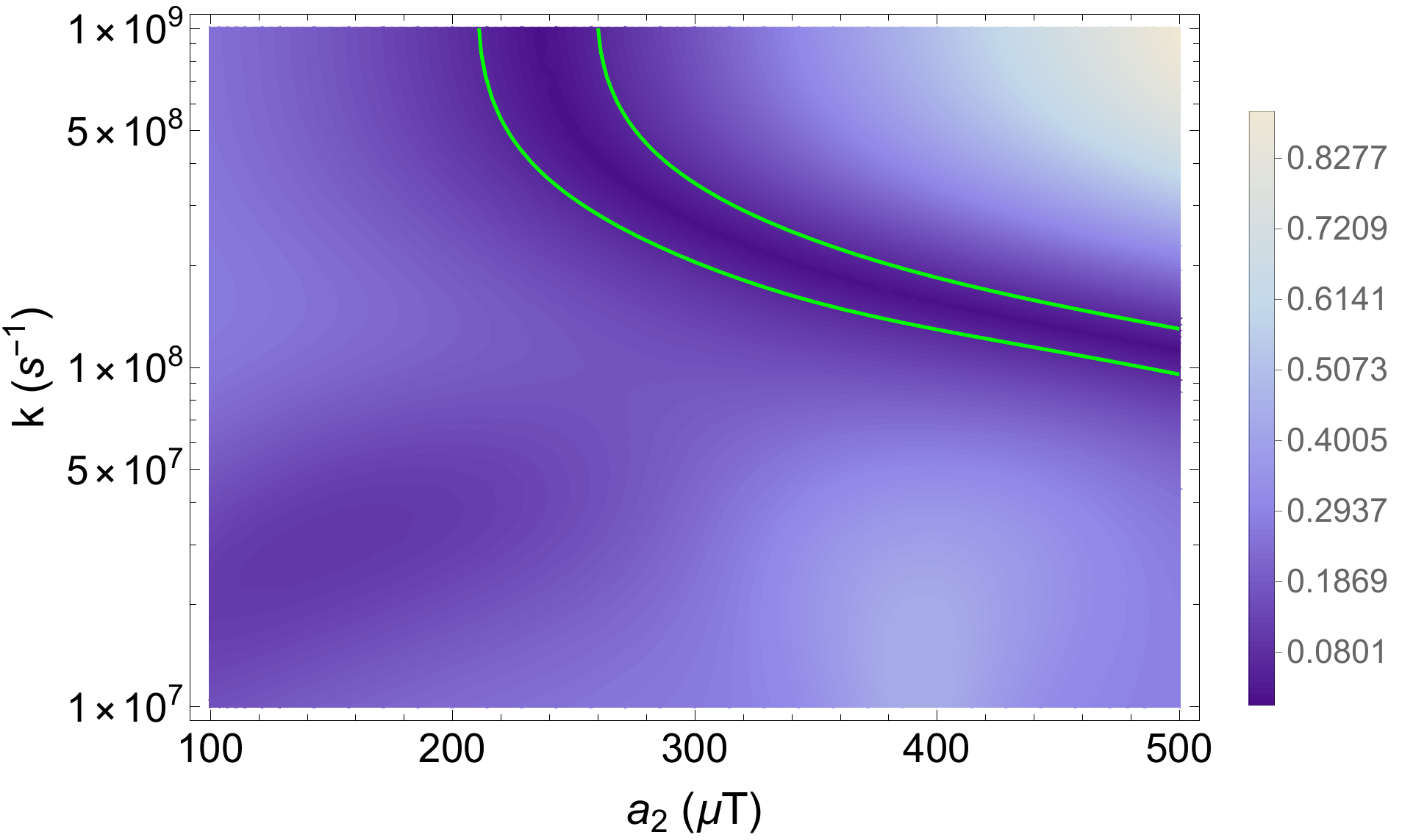}%
\caption{}%
\label{subfiga}%
\end{subfigure}\hfill%
\begin{subfigure}{.7\textwidth}
\includegraphics[width=0.7\textwidth]{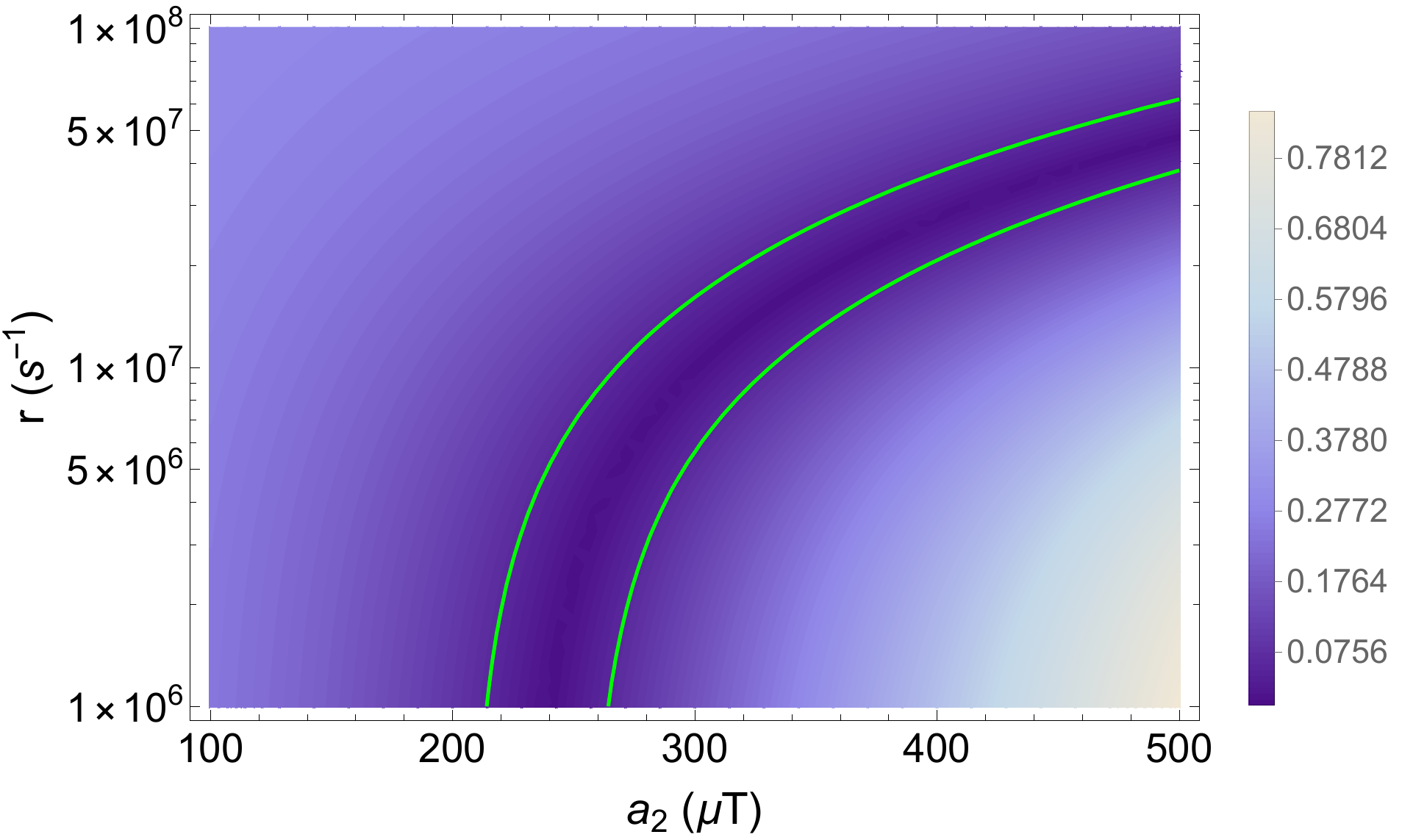}%
\caption{}%
\label{subfigb}%
\end{subfigure}\hfill%
\begin{subfigure}{.7\textwidth}
\includegraphics[width=0.7\textwidth]{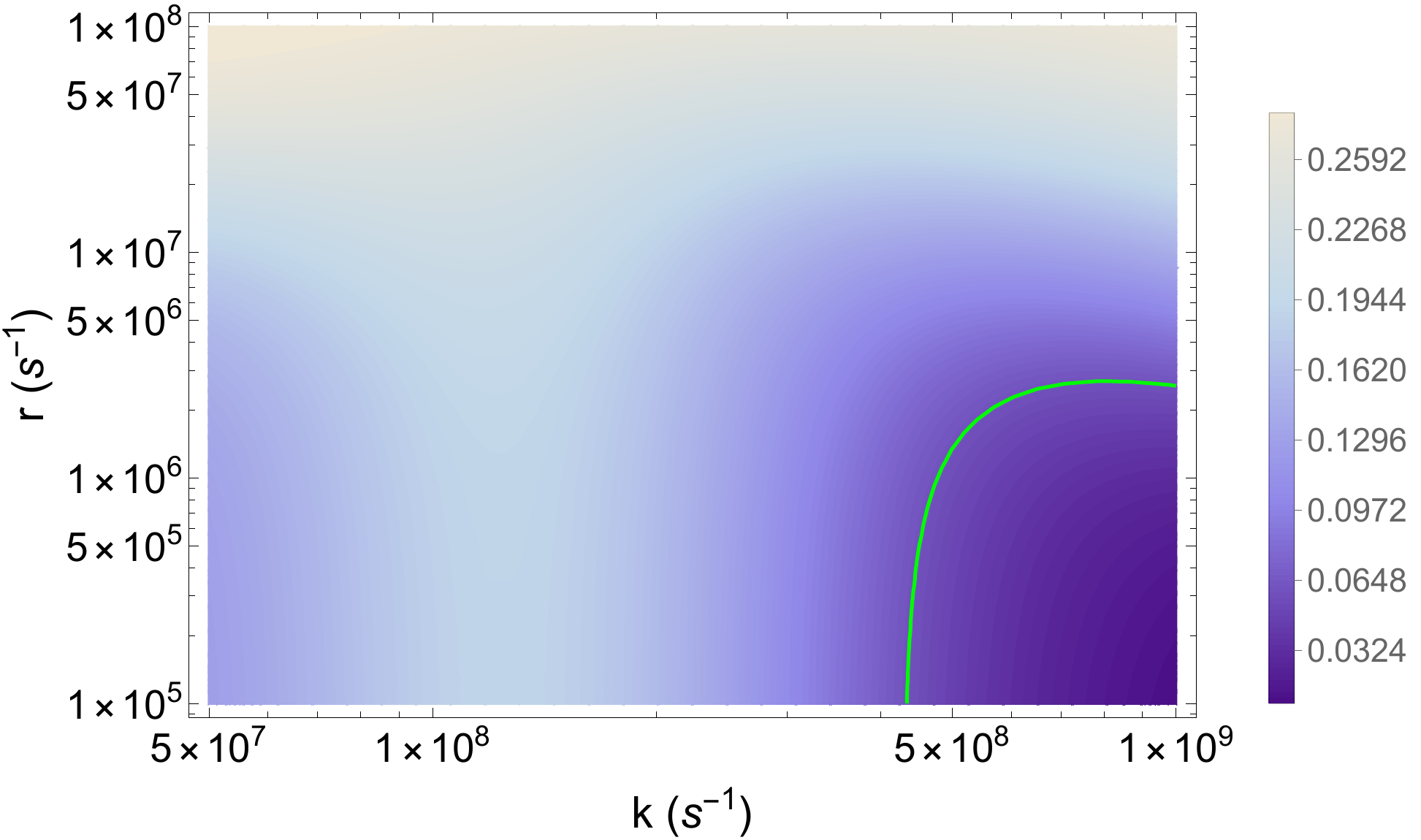}%
\caption{}%
\label{subfigc}%
\end{subfigure}%
\caption{The dependence of the agreement between the total traveled distance ratio, $TD_r$, and the triplet yield ratio, $TY_r$ of \ce{^{7}Li} over \ce{^{6}Li} on the relationship between: a) the radical pair reaction rate, $k$, and the lithium hyperfine coupling constant, $a_2$, for $r = 1.0\times10^6$ s$^{-1}$ b) the radical pair spin-coherence relaxation rate, $r$, and the lithium hyperfine coupling constant, $a_2$, for $k = 7.0\times10^8$ s$^{-1}$ and c) the RP reaction rate, $k$ and the radical pair spin-coherence relaxation rate, $r$, for $a_2 =224.4 \mu$T. In all three cases $a_1 =523.3 \mu$T and $B = 50 \mu$T. The absolute value of the difference between the prediction of radical pair mechanism, $TY_r$, and the experimental data, $TD_r$, is illustrated where the green line indicates the ranges smaller than the experimental uncertainty, $|TD_r-TY_r| \leq 0.06$.}
\label{fig:contour}
\end{figure}

\subsubsection*{Triplet Yield Calculation}
The triplet yield produced by the radical pair reaction can be obtained by tracking the spin state of the radical pair over the course of the reaction. This can be carried out by solving the Liouville-von Neumann equation, which describes the evolution of the density matrix over time. The eigenvalues and eigenvectors of the Hamiltonian can be used to determine the ultimate triplet yield ($\Phi_T$) for time periods much greater than the RP lifetime:
\begin{equation}
    \Phi_T=\frac{3k}{16(k+r)}-\frac{1}{M}\sum_{m,n=1}^{4M}\frac{\left\|\bra{m}\hat{P}^S \ket{n}\right\|^2 k(k+r)}{(k+r)^2+(\omega_m-\omega_n)^2},
\end{equation}
where $M$ is the total number of nuclear spin configurations, $\hat{P}^S$ is the singlet projection operator, $\ket{m}$ and $\ket{n}$ are eigenstates of $\hat{H}$ with corresponding eigenenergies of $\omega_m$ and $\omega_n$, respectively, $k$ is the RP lifetime rate, and $r$ is the RP spin-coherence lifetime rate.\par

\begin{figure}[ht!]
    \centering
    \includegraphics[width=0.6\textwidth]{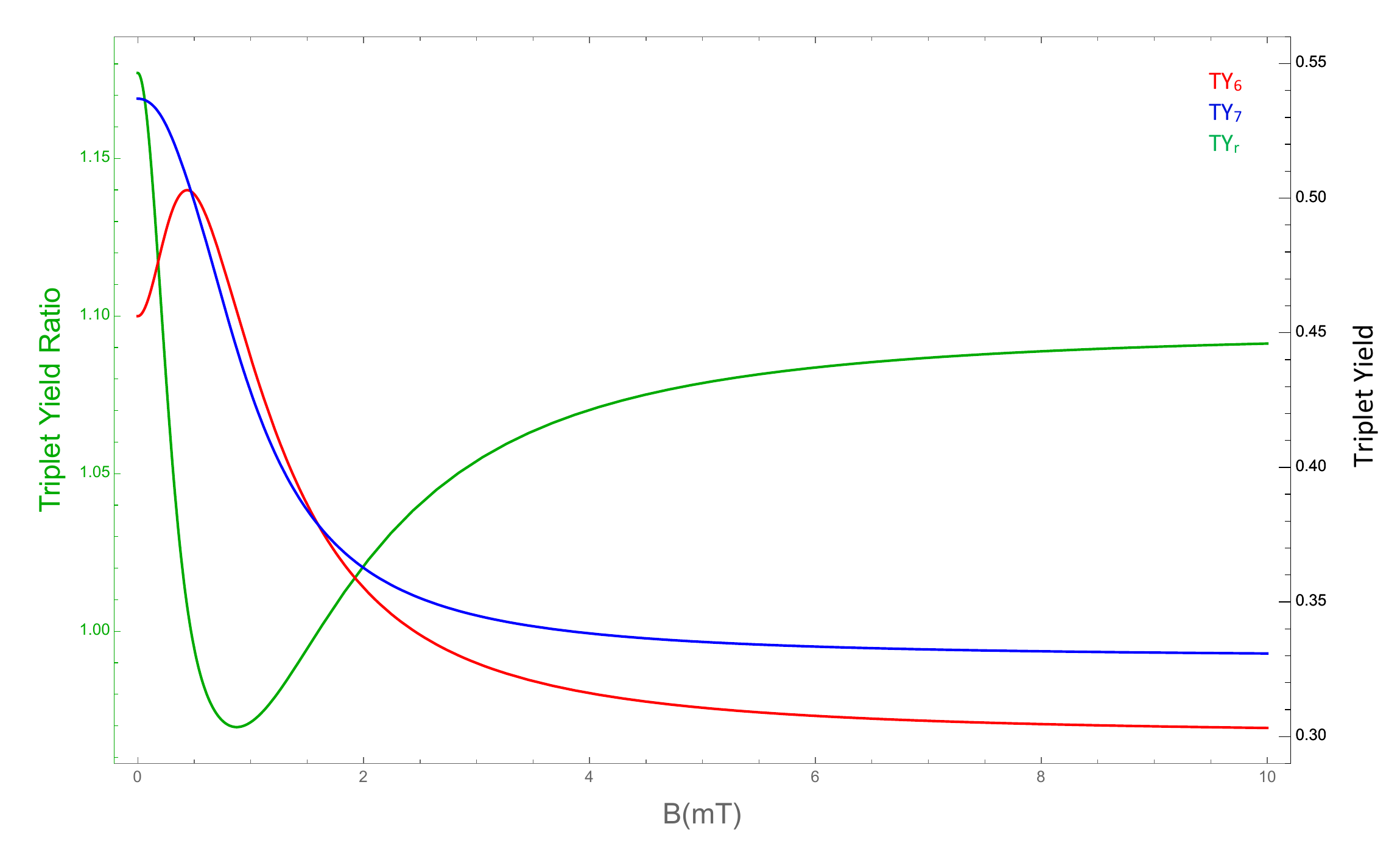}
   \caption{The dependence of the triplet yield of \ce{^{6}Li}, $TY_6$, (red) and \ce{^{7}Li}, $TY_7$, (blue) and the triplet yield ratio, $TY_r$, (green) on external magnetic field $B$ for $a_1 = 523.3 \mu$T, $a_2 = 224.4 \mu$T, $r = 1.0\times10^6$ s$^{-1}$, and $k= 4.0\times10^7$ s$^{-1}$.}
    \label{fig:tyb}
\end{figure}

Here we explore the sensitivity of the triplet yield ratio to changes in the HFCC $a_2$ between unpaired electron B and the lithium nucleus, external magnetic field strength ($B$), RP reaction rate ($k$), and RP spin-coherence relaxation rate ($r$). For the comparison between the experimental measurements and our RPM model, the absolute value of the difference between total traveled distance ratio and triplet yield, $|TD_r-TY_r|$, is presented on the $a_2$ and $k$ plane in Fig. \ref{fig:contour}a, on the $a_2$ and $r$ plane in Fig. \ref{fig:contour}b, and on the $k$ and $r$ plane in Fig. \ref{fig:contour}c. The dependence of the $TY$ for each isotope of lithium and their ratio on the external magnetic field is shown in Fig. \ref{fig:tyb}. The experimental findings of the isotopic-dependence of Li treatment for hyperactivity are reproducible for $B\in [0,200] \mu$T, which includes the geomagnetic field at different geographic locations ($25$ to $65 \mu$T) \cite{2010}. We aim to find regions in parameter space by which the triplet yield calculated from our RPM model fits with the experimental findings on the isotope dependence of the effectiveness of lithium for hyperactivity treatment. We indicate the regions within which the difference between our model and the experimental data is smaller than the uncertainty of the experimental results,$i.e.$, $|TD_r-TY_r| \leq 0.06$, as shown in Fig. \ref{fig:contour}. For a fixed external magnetic field $B=50 \mu T $ and $a_1=523.3 \mu$T, our model can predict the experimental results with quite broad ranges for the HFCC between the unpaired electron and Li nuclear spin, $a_2$, the RP reaction rate, $k$,  the RP spin relaxation rate, $r$, namely $a_2 \in [210, 500] \mu$T , $k \in [1.\times10^8, 1\times10^9]$ s$^{-1}$, and $r \in [1\times10^6,  6\times10^7]$ s$^{-1}$, see Fig. \ref{fig:contour}. The predicted values for the lithium HFCC overlaps with the range that we find for \ch{Li^{+}}-\ch{O2^{.-}} from our DFT calculations.\par

\section*{Discussion}
Our principal goal in this project was to probe whether the RPM can be the underlying mechanism behind the isotopic dependence of lithium's effectiveness for the hyperactivity treatment observed by Ettenberg \textit{et al.} \cite{Ettenberg2020}. Our results support such a mechanism. A simple radical model with a set of reasonable parameters can reproduce the experimental findings.\par

We proposed that the [\ch{FADH^{.}}... \ch{O2^{.-}}] complex is the naturally occurring RP in the circadian center of the brain and lithium interacts with the radical electron on superoxide. This is motivated by the observations \cite{abe2000lithium} that one of the possible pathways for Li's effects is Li's direct influence on the suprachiasmatic nucleus--a region in the brain containing cryptochrome protein--, that bipolar disorder is associated with imbalances in the ROS level \cite{Salim2014}, and that the lithium treatment modulates the oxidative stress level \cite{Khairova2011,MachadoVieira2007,frey2006effects,deSousa2014}. By varying the RP spin-coherence relaxation rate, RP reaction rate, and hyperfine coupling parameters our model reproduced the experimental findings of Ettenberg \textit{et al.} \cite{Ettenberg2020}. The predicted range for the lithium hyperfine coupling constant in    \ch{Li^{+}}-\ch{O2^{.-}} overlaps with our DFT calculations. \par

Let us note  that the RPM model could be adapted for other pathways by which lithium could affect BD. For example, it has also been proposed that Li could exert its effects on BD via increasing glutamate re-uptake at the N-methyl-D-aspartate (NMDA) receptor \cite{Ghasemi2011}. Inspired by the RPM's potential role in xenon-induced anesthesia \cite{Smith2021}, where there the NMDA receptor has been proposed as the target site of xenon, a similar scenario could be envisioned here. For example, oxygen might oxidize Trp in the NMDA receptor and result in the formation of [\ch{TrpH^{.-}}...\ch{O2^{.-}}] RPs, and lithium might modulate the S-T interconversion of the RPs by coupling to the unpaired electron of \ch{O2^{.-}}. \par

There remains a question on the feasibility for the \ch{O2^{.-}} radical to be involved in the RPM in this scenario due to its expected fast spin relaxation rate $r$. In the context of magnetoreception, the superoxide-based RP model has been discussed by Hogben \textit{et
al.} \cite{Hogben2009} and Player and Hore \cite{Player2019}. The authors argue that due to fast molecular rotation free \ch{O2^{.-}} has a spin relaxation lifetime on the orders of $1$ ns. The relaxation rate requirement calculated by our model yields $r$ about two orders of magnitude slower than this value, see Fig. \ref{fig:contour}. However, the same authors have also noted that this fast spin relaxation of free superoxide can be can be lowered if the molecular symmetry is reduced and the angular momentum quenched by its biological environment. Moreover, Kattnig \cite{Kattnig2017,Kattnig2017b} proposed that such fast spin relaxation of \ch{O2^{.-}} could be, in effect, reduced by the involvement of scavenger species around \ch{O2^{.-}}. For example, it has been suggested that, in the RPM, \ch{Trp^{.}} could act as a scavenger molecule for \ch{O2^{.-}} \cite{Mondal2019}. Alternatively, other RP constituents could be considered instead of \ch{O2^{.-}} to explain isotopic effects within the framework of the RPM.\par

The predicted dependence of the triple yield on changes of external magnetic field in Fig.\ref{fig:tyb} indicates that the effectiveness of the Li treatment could be enhanced by applying external magnetic fields. It would be of interest to conduct such experiments \textit{in vivo} to explore the impact of the external magnetic field on the effectiveness of the different isotopes of lithium for hyperactivity treatment. It would also be of interest, to explore isotopic nuclear-spin effects of key elements of the biological environment (particularly oxygen, but also e.g. nitrogen, carbon, and hydrogen) in experiments on the lithium effectiveness on hyperactivity.\par

In summary, our results suggest that quantum entanglement might lie behind the mechanism of lithium treatment for BD, similarly to  magnetoreception in animals \cite{Mouritsen2018} and  (as recently proposed) xenon-induced anesthesia \cite{Smith2021}.\par

This also raises the question whether the RPM could play a role in other mental disorders, and could lead to new approaches to treatment and improving efficiency of medications \cite{Raz2005}, specifically for illnesses that have been shown to be associated with oxidative stress, such as Alzheimer's \cite{Forlenza2012,Forlenza2014}, Schizophrenia\cite{Mahadik1996,Clay2010,Snyder2013}, and Parkinson's \cite{Jenner2003}. Further, it is known that light exposure affects mood and emotions \cite{Kent2009,LeGates2014,Srinivasan2006,LMcCOLL2001,Pail2011,Winkler2006}, and light is required for the formation of RPs in Crys \cite{Mller2011,Player2019}. It therefore seems that RPM could also elucidate the effects of light exposure on mood. Similarly, RPM may provide explanations for the anti-depressant effect of vitamin D \cite{Penckofer2010,Spedding2014,Parker2017} and its effects on the modulation of ROS production \cite{Uberti2016,Uberti2017}. Given that the RPM is typically associated with isotope and magnetic field effects, it would be of  great interest to search for such effects for other neurological medications. \par

Memory, learning, and subjective experience are affected by moods and emotions, and it has been proposed that cognition and subjective experience may be related to large-scale entanglement \cite{Hameroff2014,Hameroff2014b,Fisher2015,Simon2019,Adams2020}. In this highly speculative context, entangled RPs could play crucial roles as sources of entanglement, and the present work could be viewed as another piece of evidence consistent with this idea in addition to Ref. \cite{Smith2021}. In particular, superoxide radicals can give rise to singlet oxygen, which can emit photons \cite{Cifra2014}. These photons could serve as quantum messengers to establish long-distance connections \cite{Kumar2016} that might be essential for consciousness \cite{Berkovitch2020}.

\section*{Methods}
\subsection*{DFT Analysis}
The ORCA package \cite{Neese2011} was used for our \ch{Li^{+}}-\ch{O2^{.-}} DFT calculations, and the molecular structure was optimized using the dispersion-corrected PBE0 functional and def2-TZVP basis set. \par

The orbitals obtained from the optimization calculations were used to calculate orbital energies as well as the hyperfine coupling constant $a_2$. Using various DFT functionals, we obtained $a_2 \in [157.7, 282]  \mu$T, see Table \ref{tab:dft}. The calculation with the double hybrid functional \cite{Goerigk2011} RI-B2GP-PLYP with def2-QZVPP basis-set resulted $a_2 = 224.4 \mu$T, which is in the range that our RPM model predicts. We used this value in our computations. Relativistic effects were treated by a scalar relativistic Hamiltonian using the zeroth-order regular approximation (ZORA) \cite{vanLenthe1996}. Solvent effects were considered by using the conductor-like polarizable continuum model (CPCM) \cite{Marenich2009}, with a dielectric constant of 2.

\begin{table}[tbhp]
\centering
\caption{\label{tab:dft}HFCC $a_2$($\mu$T)  using different DFT fuctionals.} 
\begin{tabular}{ll}
Functional & $a_2(\mu)$T\\
\hline
BHLYP &	282\\
RI-B2GP-PLYP & 224.4\\
B3LYP &	210.8\\
B3LYP/G	& 209.8\\
PBE0 &	167.2\\
B3PW91	& 157.7\\
PW6B95	& 208.4\\
\end{tabular}
\end{table}

\section*{Data Availability}
The generated datasets and computational analysis are available from the corresponding author on reasonable request.

\bibliography{sample}

\begin{thebibliography}{10}
\urlstyle{rm}
\expandafter\ifx\csname url\endcsname\relax
  \def\url#1{\texttt{#1}}\fi
\expandafter\ifx\csname urlprefix\endcsname\relax\def\urlprefix{URL }\fi
\expandafter\ifx\csname doiprefix\endcsname\relax\def\doiprefix{DOI: }\fi
\providecommand{\bibinfo}[2]{#2}
\providecommand{\eprint}[2][]{\url{#2}}

\bibitem{MartnezArn2004}
\bibinfo{author}{Mart{\'\i}nez-Ar{\'a}n, A.} \emph{et~al.}
\newblock \bibinfo{journal}{\bibinfo{title}{Cognitive function across manic or
  hypomanic, depressed, and euthymic states in bipolar disorder}}.
\newblock {\emph{\JournalTitle{American Journal of Psychiatry}}}
  \textbf{\bibinfo{volume}{161}}, \bibinfo{pages}{262--270}
  (\bibinfo{year}{2004}).

\bibitem{Grande2016}
\bibinfo{author}{Grande, I.}, \bibinfo{author}{Berk, M.},
  \bibinfo{author}{Birmaher, B.} \& \bibinfo{author}{Vieta, E.}
\newblock \bibinfo{journal}{\bibinfo{title}{Bipolar disorder}}.
\newblock {\emph{\JournalTitle{The Lancet}}} \textbf{\bibinfo{volume}{387}},
  \bibinfo{pages}{1561--1572} (\bibinfo{year}{2016}).

\bibitem{Geddes2013}
\bibinfo{author}{Geddes, J.~R.} \& \bibinfo{author}{Miklowitz, D.~J.}
\newblock \bibinfo{journal}{\bibinfo{title}{Treatment of bipolar disorder}}.
\newblock {\emph{\JournalTitle{The lancet}}} \textbf{\bibinfo{volume}{381}},
  \bibinfo{pages}{1672--1682} (\bibinfo{year}{2013}).

\bibitem{McIntyre2020}
\bibinfo{author}{McIntyre, R.~S.} \emph{et~al.}
\newblock \bibinfo{journal}{\bibinfo{title}{Bipolar disorders}}.
\newblock {\emph{\JournalTitle{The Lancet}}} \textbf{\bibinfo{volume}{396}},
  \bibinfo{pages}{1841--1856} (\bibinfo{year}{2020}).

\bibitem{Malhi2017}
\bibinfo{author}{Malhi, G.~S.}, \bibinfo{author}{Gessler, D.} \&
  \bibinfo{author}{Outhred, T.}
\newblock \bibinfo{journal}{\bibinfo{title}{The use of lithium for the
  treatment of bipolar disorder: recommendations from clinical practice
  guidelines}}.
\newblock {\emph{\JournalTitle{Journal of affective disorders}}}
  \textbf{\bibinfo{volume}{217}}, \bibinfo{pages}{266--280}
  (\bibinfo{year}{2017}).

\bibitem{MachadoVieira2009}
\bibinfo{author}{Machado-Vieira, R.}, \bibinfo{author}{Manji, H.~K.} \&
  \bibinfo{author}{Zarate~Jr, C.~A.}
\newblock \bibinfo{journal}{\bibinfo{title}{The role of lithium in the
  treatment of bipolar disorder: convergent evidence for neurotrophic effects
  as a unifying hypothesis}}.
\newblock {\emph{\JournalTitle{Bipolar disorders}}}
  \textbf{\bibinfo{volume}{11}}, \bibinfo{pages}{92--109}
  (\bibinfo{year}{2009}).

\bibitem{Khairova2011}
\bibinfo{author}{Khairova, R.} \emph{et~al.}
\newblock \bibinfo{journal}{\bibinfo{title}{Effects of lithium on oxidative
  stress parameters in healthy subjects}}.
\newblock {\emph{\JournalTitle{Molecular Medicine Reports}}}
  \textbf{\bibinfo{volume}{5}}, \bibinfo{pages}{680--682}
  (\bibinfo{year}{2012}).

\bibitem{Roux2017}
\bibinfo{author}{Roux, M.} \& \bibinfo{author}{Dosseto, A.}
\newblock \bibinfo{journal}{\bibinfo{title}{From direct to indirect lithium
  targets: a comprehensive review of omics data}}.
\newblock {\emph{\JournalTitle{Metallomics}}} \textbf{\bibinfo{volume}{9}},
  \bibinfo{pages}{1326--1351} (\bibinfo{year}{2017}).

\bibitem{Harwood2004}
\bibinfo{author}{Harwood, A.}
\newblock \bibinfo{journal}{\bibinfo{title}{Lithium and bipolar mood disorder:
  the inositol-depletion hypothesis revisited}}.
\newblock {\emph{\JournalTitle{Molecular psychiatry}}}
  \textbf{\bibinfo{volume}{10}}, \bibinfo{pages}{117--126}
  (\bibinfo{year}{2005}).

\bibitem{Sechzer1986}
\bibinfo{author}{Sechzer, J.~A.}, \bibinfo{author}{Lieberman, K.~W.},
  \bibinfo{author}{Alexander, G.~J.}, \bibinfo{author}{Weidman, D.} \&
  \bibinfo{author}{Stokes, P.~E.}
\newblock \bibinfo{journal}{\bibinfo{title}{Aberrant parenting and delayed
  offspring development in rats exposed to lithium}}.
\newblock {\emph{\JournalTitle{Biological psychiatry}}}
  \textbf{\bibinfo{volume}{21}}, \bibinfo{pages}{1258--1266}
  (\bibinfo{year}{1986}).

\bibitem{Ettenberg2020}
\bibinfo{author}{Ettenberg, A.} \emph{et~al.}
\newblock \bibinfo{journal}{\bibinfo{title}{Differential effects of lithium
  isotopes in a ketamine-induced hyperactivity model of mania}}.
\newblock {\emph{\JournalTitle{Pharmacology Biochemistry and Behavior}}}
  \textbf{\bibinfo{volume}{190}}, \bibinfo{pages}{172875}
  (\bibinfo{year}{2020}).

\bibitem{Andreazza2008}
\bibinfo{author}{Andreazza, A.~C.} \emph{et~al.}
\newblock \bibinfo{journal}{\bibinfo{title}{Oxidative stress markers in bipolar
  disorder: a meta-analysis}}.
\newblock {\emph{\JournalTitle{Journal of affective disorders}}}
  \textbf{\bibinfo{volume}{111}}, \bibinfo{pages}{135--144}
  (\bibinfo{year}{2008}).

\bibitem{Yumru2009}
\bibinfo{author}{Yumru, M.} \emph{et~al.}
\newblock \bibinfo{journal}{\bibinfo{title}{Oxidative imbalance in bipolar
  disorder subtypes: a comparative study}}.
\newblock {\emph{\JournalTitle{Progress in Neuro-Psychopharmacology and
  Biological Psychiatry}}} \textbf{\bibinfo{volume}{33}},
  \bibinfo{pages}{1070--1074} (\bibinfo{year}{2009}).

\bibitem{Salim2014}
\bibinfo{author}{Salim, S.}
\newblock \bibinfo{journal}{\bibinfo{title}{Oxidative stress and psychological
  disorders}}.
\newblock {\emph{\JournalTitle{Current neuropharmacology}}}
  \textbf{\bibinfo{volume}{12}}, \bibinfo{pages}{140--147}
  (\bibinfo{year}{2014}).

\bibitem{MachadoVieira2007}
\bibinfo{author}{Machado-Vieira, R.} \emph{et~al.}
\newblock \bibinfo{journal}{\bibinfo{title}{Oxidative stress parameters in
  unmedicated and treated bipolar subjects during initial manic episode: a
  possible role for lithium antioxidant effects}}.
\newblock {\emph{\JournalTitle{Neuroscience letters}}}
  \textbf{\bibinfo{volume}{421}}, \bibinfo{pages}{33--36}
  (\bibinfo{year}{2007}).

\bibitem{Ng2008}
\bibinfo{author}{Ng, F.}, \bibinfo{author}{Berk, M.}, \bibinfo{author}{Dean,
  O.} \& \bibinfo{author}{Bush, A.~I.}
\newblock \bibinfo{journal}{\bibinfo{title}{Oxidative stress in psychiatric
  disorders: evidence base and therapeutic implications}}.
\newblock {\emph{\JournalTitle{International Journal of
  Neuropsychopharmacology}}} \textbf{\bibinfo{volume}{11}},
  \bibinfo{pages}{851--876} (\bibinfo{year}{2008}).

\bibitem{Lee2013}
\bibinfo{author}{Lee, S.-Y.} \emph{et~al.}
\newblock \bibinfo{journal}{\bibinfo{title}{Oxidative/nitrosative stress and
  antidepressants: targets for novel antidepressants}}.
\newblock {\emph{\JournalTitle{Progress in Neuro-Psychopharmacology and
  Biological Psychiatry}}} \textbf{\bibinfo{volume}{46}},
  \bibinfo{pages}{224--235} (\bibinfo{year}{2013}).

\bibitem{Brown2014}
\bibinfo{author}{Brown, N.~C.}, \bibinfo{author}{Andreazza, A.~C.} \&
  \bibinfo{author}{Young, L.~T.}
\newblock \bibinfo{journal}{\bibinfo{title}{An updated meta-analysis of
  oxidative stress markers in bipolar disorder}}.
\newblock {\emph{\JournalTitle{Psychiatry Research}}}
  \textbf{\bibinfo{volume}{218}}, \bibinfo{pages}{61--68}
  (\bibinfo{year}{2014}).

\bibitem{Berk2011}
\bibinfo{author}{Berk, M.} \emph{et~al.}
\newblock \bibinfo{journal}{\bibinfo{title}{Pathways underlying
  neuroprogression in bipolar disorder: focus on inflammation, oxidative stress
  and neurotrophic factors}}.
\newblock {\emph{\JournalTitle{Neuroscience \& biobehavioral reviews}}}
  \textbf{\bibinfo{volume}{35}}, \bibinfo{pages}{804--817}
  (\bibinfo{year}{2011}).

\bibitem{frey2006effects}
\bibinfo{author}{Frey, B.~N.} \emph{et~al.}
\newblock \bibinfo{journal}{\bibinfo{title}{Effects of lithium and valproate on
  amphetamine-induced oxidative stress generation in an animal model of
  mania}}.
\newblock {\emph{\JournalTitle{Journal of Psychiatry and Neuroscience}}}
  \textbf{\bibinfo{volume}{31}}, \bibinfo{pages}{326} (\bibinfo{year}{2006}).

\bibitem{deSousa2014}
\bibinfo{author}{de~Sousa, R.~T.} \emph{et~al.}
\newblock \bibinfo{journal}{\bibinfo{title}{Oxidative stress in early stage
  bipolar disorder and the association with response to lithium}}.
\newblock {\emph{\JournalTitle{Journal of psychiatric research}}}
  \textbf{\bibinfo{volume}{50}}, \bibinfo{pages}{36--41}
  (\bibinfo{year}{2014}).

\bibitem{Sies2017}
\bibinfo{author}{Sies, H.}, \bibinfo{author}{Berndt, C.} \&
  \bibinfo{author}{Jones, D.~P.}
\newblock \bibinfo{journal}{\bibinfo{title}{Oxidative stress}}.
\newblock {\emph{\JournalTitle{Annual review of biochemistry}}}
  \textbf{\bibinfo{volume}{86}}, \bibinfo{pages}{715--748}
  (\bibinfo{year}{2017}).

\bibitem{Steiner1989}
\bibinfo{author}{Steiner, U.~E.} \& \bibinfo{author}{Ulrich, T.}
\newblock \bibinfo{journal}{\bibinfo{title}{Magnetic field effects in chemical
  kinetics and related phenomena}}.
\newblock {\emph{\JournalTitle{Chemical Reviews}}}
  \textbf{\bibinfo{volume}{89}}, \bibinfo{pages}{51--147}
  (\bibinfo{year}{1989}).

\bibitem{Hayashi2004}
\bibinfo{author}{Hayashi, H.}
\newblock \emph{\bibinfo{title}{Introduction to dynamic spin chemistry:
  magnetic field effects on chemical and biochemical reactions}},
  vol.~\bibinfo{volume}{8} (\bibinfo{publisher}{World Scientific Publishing
  Company}, \bibinfo{year}{2004}).

\bibitem{Schulten1976}
\bibinfo{author}{Schulten, K.}, \bibinfo{author}{Staerk, H.},
  \bibinfo{author}{Weller, A.}, \bibinfo{author}{Werner, H.-J.} \&
  \bibinfo{author}{Nickel, B.}
\newblock \bibinfo{journal}{\bibinfo{title}{Magnetic field dependence of the
  geminate recombination of radical ion pairs in polar solvents}}.
\newblock {\emph{\JournalTitle{Z. Phys. Chem}}} \textbf{\bibinfo{volume}{101}},
  \bibinfo{pages}{371--390} (\bibinfo{year}{1976}).

\bibitem{Brocklehurst1974}
\bibinfo{author}{Brocklehurst, B.} \emph{et~al.}
\newblock \bibinfo{journal}{\bibinfo{title}{The effect of a magnetic field on
  the singlet/triplet ratio in geminate ion recombination}}.
\newblock {\emph{\JournalTitle{Chemical Physics Letters}}}
  \textbf{\bibinfo{volume}{28}}, \bibinfo{pages}{361--363}
  (\bibinfo{year}{1974}).

\bibitem{Jones2016}
\bibinfo{author}{Jones, A.~R.}
\newblock \bibinfo{journal}{\bibinfo{title}{Magnetic field effects in
  proteins}}.
\newblock {\emph{\JournalTitle{Molecular Physics}}}
  \textbf{\bibinfo{volume}{114}}, \bibinfo{pages}{1691--1702}
  (\bibinfo{year}{2016}).

\bibitem{Timmel1998}
\bibinfo{author}{Timmel, C.~R.}, \bibinfo{author}{Till, U.},
  \bibinfo{author}{Brocklehurst, B.}, \bibinfo{author}{Mclauchlan, K.~A.} \&
  \bibinfo{author}{Hore, P.~J.}
\newblock \bibinfo{journal}{\bibinfo{title}{Effects of weak magnetic fields on
  free radical recombination reactions}}.
\newblock {\emph{\JournalTitle{Molecular Physics}}}
  \textbf{\bibinfo{volume}{95}}, \bibinfo{pages}{71--89}
  (\bibinfo{year}{1998}).

\bibitem{Ball2011}
\bibinfo{author}{Ball, P.}
\newblock \bibinfo{journal}{\bibinfo{title}{Physics of life: The dawn of
  quantum biology}}.
\newblock {\emph{\JournalTitle{Nature News}}} \textbf{\bibinfo{volume}{474}},
  \bibinfo{pages}{272--274} (\bibinfo{year}{2011}).

\bibitem{mcfadden2016life}
\bibinfo{author}{McFadden, J.} \& \bibinfo{author}{Al-Khalili, J.}
\newblock \emph{\bibinfo{title}{Life on the edge: the coming of age of quantum
  biology}} (\bibinfo{publisher}{Crown Publishing Group (NY)},
  \bibinfo{year}{2016}).

\bibitem{Kim2021}
\bibinfo{author}{Kim, Y.} \emph{et~al.}
\newblock \bibinfo{journal}{\bibinfo{title}{Quantum biology: An update and
  perspective}}.
\newblock {\emph{\JournalTitle{Quantum Reports}}} \textbf{\bibinfo{volume}{3}},
  \bibinfo{pages}{80--126} (\bibinfo{year}{2021}).

\bibitem{Hore2016}
\bibinfo{author}{Hore, P.~J.} \& \bibinfo{author}{Mouritsen, H.}
\newblock \bibinfo{journal}{\bibinfo{title}{The radical-pair mechanism of
  magnetoreception}}.
\newblock {\emph{\JournalTitle{Annual review of biophysics}}}
  \textbf{\bibinfo{volume}{45}}, \bibinfo{pages}{299--344}
  (\bibinfo{year}{2016}).

\bibitem{Mouritsen2018}
\bibinfo{author}{Mouritsen, H.}
\newblock \bibinfo{journal}{\bibinfo{title}{Long-distance navigation and
  magnetoreception in migratory animals}}.
\newblock {\emph{\JournalTitle{Nature}}} \textbf{\bibinfo{volume}{558}},
  \bibinfo{pages}{50--59} (\bibinfo{year}{2018}).

\bibitem{Hore2020}
\bibinfo{author}{Hore, P.~J.}, \bibinfo{author}{Ivanov, K.~L.} \&
  \bibinfo{author}{Wasielewski, M.~R.}
\newblock \bibinfo{journal}{\bibinfo{title}{Spin chemistry}}.
\newblock {\emph{\JournalTitle{The Journal of Chemical Physics}}}
  \textbf{\bibinfo{volume}{152}}, \bibinfo{pages}{120401}
  (\bibinfo{year}{2020}).

\bibitem{Smith2021}
\bibinfo{author}{Smith, J.}, \bibinfo{author}{Zadeh-Haghighi, H.},
  \bibinfo{author}{Salahub, D.} \& \bibinfo{author}{Simon, C.}
\newblock \bibinfo{journal}{\bibinfo{title}{Radical pairs may play a role in
  xenon-induced general anesthesia}}.
\newblock {\emph{\JournalTitle{Scientific Reports}}}
  \textbf{\bibinfo{volume}{11}}, \bibinfo{pages}{6287} (\bibinfo{year}{2021}).

\bibitem{Ritz2000}
\bibinfo{author}{Ritz, T.}, \bibinfo{author}{Adem, S.} \&
  \bibinfo{author}{Schulten, K.}
\newblock \bibinfo{journal}{\bibinfo{title}{A model for photoreceptor-based
  magnetoreception in birds}}.
\newblock {\emph{\JournalTitle{Biophysical journal}}}
  \textbf{\bibinfo{volume}{78}}, \bibinfo{pages}{707--718}
  (\bibinfo{year}{2000}).

\bibitem{Giovani2003}
\bibinfo{author}{Giovani, B.}, \bibinfo{author}{Byrdin, M.},
  \bibinfo{author}{Ahmad, M.} \& \bibinfo{author}{Brettel, K.}
\newblock \bibinfo{journal}{\bibinfo{title}{Light-induced electron transfer in
  a cryptochrome blue-light photoreceptor}}.
\newblock {\emph{\JournalTitle{Nature Structural \& Molecular Biology}}}
  \textbf{\bibinfo{volume}{10}}, \bibinfo{pages}{489--490}
  (\bibinfo{year}{2003}).

\bibitem{Hong2020}
\bibinfo{author}{Hong, G.}, \bibinfo{author}{Pachter, R.},
  \bibinfo{author}{Essen, L.-O.} \& \bibinfo{author}{Ritz, T.}
\newblock \bibinfo{journal}{\bibinfo{title}{Electron transfer and spin dynamics
  of the radical-pair in the cryptochrome from chlamydomonas reinhardtii by
  computational analysis}}.
\newblock {\emph{\JournalTitle{The Journal of chemical physics}}}
  \textbf{\bibinfo{volume}{152}}, \bibinfo{pages}{065101}
  (\bibinfo{year}{2020}).

\bibitem{Mller2011}
\bibinfo{author}{M{\"u}ller, P.} \& \bibinfo{author}{Ahmad, M.}
\newblock \bibinfo{journal}{\bibinfo{title}{Light-activated cryptochrome reacts
  with molecular oxygen to form a flavin--superoxide radical pair consistent
  with magnetoreception}}.
\newblock {\emph{\JournalTitle{Journal of Biological Chemistry}}}
  \textbf{\bibinfo{volume}{286}}, \bibinfo{pages}{21033--21040}
  (\bibinfo{year}{2011}).

\bibitem{Romero2018}
\bibinfo{author}{Romero, E.}, \bibinfo{author}{Gómez~Castellanos, J.~R.},
  \bibinfo{author}{Gadda, G.}, \bibinfo{author}{Fraaije, M.~W.} \&
  \bibinfo{author}{Mattevi, A.}
\newblock \bibinfo{journal}{\bibinfo{title}{Same substrate, many reactions:
  Oxygen activation in flavoenzymes}}.
\newblock {\emph{\JournalTitle{Chemical reviews}}}
  \textbf{\bibinfo{volume}{118}}, \bibinfo{pages}{1742--1769}
  (\bibinfo{year}{2018}).

\bibitem{Chaiyen2012}
\bibinfo{author}{Chaiyen, P.}, \bibinfo{author}{Fraaije, M.~W.} \&
  \bibinfo{author}{Mattevi, A.}
\newblock \bibinfo{journal}{\bibinfo{title}{The enigmatic reaction of flavins
  with oxygen}}.
\newblock {\emph{\JournalTitle{Trends in biochemical sciences}}}
  \textbf{\bibinfo{volume}{37}}, \bibinfo{pages}{373--380}
  (\bibinfo{year}{2012}).

\bibitem{Mondal2019}
\bibinfo{author}{Mondal, P.} \& \bibinfo{author}{Huix-Rotllant, M.}
\newblock \bibinfo{journal}{\bibinfo{title}{Theoretical insights into the
  formation and stability of radical oxygen species in cryptochromes}}.
\newblock {\emph{\JournalTitle{Physical Chemistry Chemical Physics}}}
  \textbf{\bibinfo{volume}{21}}, \bibinfo{pages}{8874--8882}
  (\bibinfo{year}{2019}).

\bibitem{Wiltschko2016}
\bibinfo{author}{Wiltschko, R.}, \bibinfo{author}{Ahmad, M.},
  \bibinfo{author}{Nie{\ss}ner, C.}, \bibinfo{author}{Gehring, D.} \&
  \bibinfo{author}{Wiltschko, W.}
\newblock \bibinfo{journal}{\bibinfo{title}{Light-dependent magnetoreception in
  birds: the crucial step occurs in the dark}}.
\newblock {\emph{\JournalTitle{Journal of The Royal Society Interface}}}
  \textbf{\bibinfo{volume}{13}}, \bibinfo{pages}{20151010}
  (\bibinfo{year}{2016}).

\bibitem{Hogben2009}
\bibinfo{author}{Hogben, H.~J.}, \bibinfo{author}{Efimova, O.},
  \bibinfo{author}{Wagner-Rundell, N.}, \bibinfo{author}{Timmel, C.~R.} \&
  \bibinfo{author}{Hore, P.}
\newblock \bibinfo{journal}{\bibinfo{title}{Possible involvement of superoxide
  and dioxygen with cryptochrome in avian magnetoreception: origin of zeeman
  resonances observed by in vivo epr spectroscopy}}.
\newblock {\emph{\JournalTitle{Chemical Physics Letters}}}
  \textbf{\bibinfo{volume}{480}}, \bibinfo{pages}{118--122}
  (\bibinfo{year}{2009}).

\bibitem{Zhang2021}
\bibinfo{author}{Zhang, B.} \emph{et~al.}
\newblock \bibinfo{journal}{\bibinfo{title}{Long-term exposure to a
  hypomagnetic field attenuates adult hippocampal neurogenesis and cognition}}.
\newblock {\emph{\JournalTitle{Nature communications}}}
  \textbf{\bibinfo{volume}{12}}, \bibinfo{pages}{1--17} (\bibinfo{year}{2021}).

\bibitem{Usselman2016}
\bibinfo{author}{Usselman, R.~J.} \emph{et~al.}
\newblock \bibinfo{journal}{\bibinfo{title}{The quantum biology of reactive
  oxygen species partitioning impacts cellular bioenergetics}}.
\newblock {\emph{\JournalTitle{Scientific reports}}}
  \textbf{\bibinfo{volume}{6}}, \bibinfo{pages}{1--6} (\bibinfo{year}{2016}).

\bibitem{Abreu2015}
\bibinfo{author}{Abreu, T.} \& \bibinfo{author}{Bragan{\c{c}}a, M.}
\newblock \bibinfo{journal}{\bibinfo{title}{The bipolarity of light and dark: a
  review on bipolar disorder and circadian cycles}}.
\newblock {\emph{\JournalTitle{Journal of affective disorders}}}
  \textbf{\bibinfo{volume}{185}}, \bibinfo{pages}{219--229}
  (\bibinfo{year}{2015}).

\bibitem{Takahashi2008}
\bibinfo{author}{Takahashi, J.~S.}, \bibinfo{author}{Hong, H.-K.},
  \bibinfo{author}{Ko, C.~H.} \& \bibinfo{author}{McDearmon, E.~L.}
\newblock \bibinfo{journal}{\bibinfo{title}{The genetics of mammalian circadian
  order and disorder: implications for physiology and disease}}.
\newblock {\emph{\JournalTitle{Nature reviews genetics}}}
  \textbf{\bibinfo{volume}{9}}, \bibinfo{pages}{764--775}
  (\bibinfo{year}{2008}).

\bibitem{Yin2006}
\bibinfo{author}{Yin, L.}, \bibinfo{author}{Wang, J.}, \bibinfo{author}{Klein,
  P.~S.} \& \bibinfo{author}{Lazar, M.~A.}
\newblock \bibinfo{journal}{\bibinfo{title}{Nuclear receptor rev-erb$\alpha$ is
  a critical lithium-sensitive component of the circadian clock}}.
\newblock {\emph{\JournalTitle{Science}}} \textbf{\bibinfo{volume}{311}},
  \bibinfo{pages}{1002--1005} (\bibinfo{year}{2006}).

\bibitem{Li2012}
\bibinfo{author}{Li, J.}, \bibinfo{author}{Lu, W.-Q.},
  \bibinfo{author}{Beesley, S.}, \bibinfo{author}{Loudon, A.~S.} \&
  \bibinfo{author}{Meng, Q.-J.}
\newblock \bibinfo{journal}{\bibinfo{title}{Lithium impacts on the amplitude
  and period of the molecular circadian clockwork}}.
\newblock {\emph{\JournalTitle{PloS one}}} \textbf{\bibinfo{volume}{7}},
  \bibinfo{pages}{e33292} (\bibinfo{year}{2012}).

\bibitem{abe2000lithium}
\bibinfo{author}{Abe, M.}, \bibinfo{author}{Herzog, E.~D.} \&
  \bibinfo{author}{Block, G.~D.}
\newblock \bibinfo{journal}{\bibinfo{title}{Lithium lengthens the circadian
  period of individual suprachiasmatic nucleus neurons}}.
\newblock {\emph{\JournalTitle{Neuroreport}}} \textbf{\bibinfo{volume}{11}},
  \bibinfo{pages}{3261--3264} (\bibinfo{year}{2000}).

\bibitem{Osland2010}
\bibinfo{author}{Osland, T.~M.} \emph{et~al.}
\newblock \bibinfo{journal}{\bibinfo{title}{Lithium differentially affects
  clock gene expression in serum-shocked nih-3t3 cells}}.
\newblock {\emph{\JournalTitle{Journal of psychopharmacology}}}
  \textbf{\bibinfo{volume}{25}}, \bibinfo{pages}{924--933}
  (\bibinfo{year}{2011}).

\bibitem{Welsh2010}
\bibinfo{author}{Welsh, D.~K.}, \bibinfo{author}{Takahashi, J.~S.} \&
  \bibinfo{author}{Kay, S.~A.}
\newblock \bibinfo{journal}{\bibinfo{title}{Suprachiasmatic nucleus: cell
  autonomy and network properties}}.
\newblock {\emph{\JournalTitle{Annual review of physiology}}}
  \textbf{\bibinfo{volume}{72}}, \bibinfo{pages}{551--577}
  (\bibinfo{year}{2010}).

\bibitem{Lavebratt2010}
\bibinfo{author}{Lavebratt, C.} \emph{et~al.}
\newblock \bibinfo{journal}{\bibinfo{title}{Cry2 is associated with
  depression}}.
\newblock {\emph{\JournalTitle{Plos one}}} \textbf{\bibinfo{volume}{5}},
  \bibinfo{pages}{e9407} (\bibinfo{year}{2010}).

\bibitem{Dufor2019}
\bibinfo{author}{Dufor, T.} \emph{et~al.}
\newblock \bibinfo{journal}{\bibinfo{title}{Neural circuit repair by
  low-intensity magnetic stimulation requires cellular magnetoreceptors and
  specific stimulation patterns}}.
\newblock {\emph{\JournalTitle{Science advances}}}
  \textbf{\bibinfo{volume}{5}}, \bibinfo{pages}{eaav9847}
  (\bibinfo{year}{2019}).

\bibitem{Fisher2015}
\bibinfo{author}{Fisher, M.~P.}
\newblock \bibinfo{journal}{\bibinfo{title}{Quantum cognition: The possibility
  of processing with nuclear spins in the brain}}.
\newblock {\emph{\JournalTitle{Annals of Physics}}}
  \textbf{\bibinfo{volume}{362}}, \bibinfo{pages}{593--602}
  (\bibinfo{year}{2015}).

\bibitem{Swift2018}
\bibinfo{author}{Swift, M.~W.}, \bibinfo{author}{Van~de Walle, C.~G.} \&
  \bibinfo{author}{Fisher, M.~P.}
\newblock \bibinfo{journal}{\bibinfo{title}{Posner molecules: from atomic
  structure to nuclear spins}}.
\newblock {\emph{\JournalTitle{Physical Chemistry Chemical Physics}}}
  \textbf{\bibinfo{volume}{20}}, \bibinfo{pages}{12373--12380}
  (\bibinfo{year}{2018}).

\bibitem{Chen2020}
\bibinfo{author}{Chen, R.}, \bibinfo{author}{Li, N.}, \bibinfo{author}{Qian,
  H.}, \bibinfo{author}{Zhao, R.-H.} \& \bibinfo{author}{Zhang, S.-H.}
\newblock \bibinfo{journal}{\bibinfo{title}{Experimental evidence refuting the
  assumption of phosphorus-31 nuclear-spin entanglement-mediated
  consciousness}}.
\newblock {\emph{\JournalTitle{Journal of Integrative Neuroscience}}}
  \textbf{\bibinfo{volume}{19}}, \bibinfo{pages}{595--600}
  (\bibinfo{year}{2020}).

\bibitem{Efimova2008}
\bibinfo{author}{Efimova, O.} \& \bibinfo{author}{Hore, P.}
\newblock \bibinfo{journal}{\bibinfo{title}{Role of exchange and dipolar
  interactions in the radical pair model of the avian magnetic compass}}.
\newblock {\emph{\JournalTitle{Biophysical Journal}}}
  \textbf{\bibinfo{volume}{94}}, \bibinfo{pages}{1565--1574}
  (\bibinfo{year}{2008}).

\bibitem{Maeda2012}
\bibinfo{author}{Maeda, K.} \emph{et~al.}
\newblock \bibinfo{journal}{\bibinfo{title}{Magnetically sensitive
  light-induced reactions in cryptochrome are consistent with its proposed role
  as a magnetoreceptor}}.
\newblock {\emph{\JournalTitle{Proceedings of the National Academy of
  Sciences}}} \textbf{\bibinfo{volume}{109}}, \bibinfo{pages}{4774--4779}
  (\bibinfo{year}{2012}).

\bibitem{Hore2019}
\bibinfo{author}{Hore, P.~J.}
\newblock \bibinfo{journal}{\bibinfo{title}{Upper bound on the biological
  effects of 50/60 hz magnetic fields mediated by radical pairs}}.
\newblock {\emph{\JournalTitle{Elife}}} \textbf{\bibinfo{volume}{8}},
  \bibinfo{pages}{e44179} (\bibinfo{year}{2019}).

\bibitem{Lau2011}
\bibinfo{author}{Lau, K.~C.}, \bibinfo{author}{Curtiss, L.~A.} \&
  \bibinfo{author}{Greeley, J.}
\newblock \bibinfo{journal}{\bibinfo{title}{Density functional investigation of
  the thermodynamic stability of lithium oxide bulk crystalline structures as a
  function of oxygen pressure}}.
\newblock {\emph{\JournalTitle{The Journal of Physical Chemistry C}}}
  \textbf{\bibinfo{volume}{115}}, \bibinfo{pages}{23625--23633}
  (\bibinfo{year}{2011}).

\bibitem{2010}
\bibinfo{author}{Finlay, C.~C.} \emph{et~al.}
\newblock \bibinfo{journal}{\bibinfo{title}{International geomagnetic reference
  field: the eleventh generation}}.
\newblock {\emph{\JournalTitle{Geophysical Journal International}}}
  \textbf{\bibinfo{volume}{183}}, \bibinfo{pages}{1216--1230}
  (\bibinfo{year}{2010}).

\bibitem{Ghasemi2011}
\bibinfo{author}{Ghasemi, M.} \& \bibinfo{author}{Dehpour, A.~R.}
\newblock \bibinfo{journal}{\bibinfo{title}{The nmda receptor/nitric oxide
  pathway: a target for the therapeutic and toxic effects of lithium}}.
\newblock {\emph{\JournalTitle{Trends in pharmacological sciences}}}
  \textbf{\bibinfo{volume}{32}}, \bibinfo{pages}{420--434}
  (\bibinfo{year}{2011}).

\bibitem{Player2019}
\bibinfo{author}{Player, T.~C.} \& \bibinfo{author}{Hore, P.}
\newblock \bibinfo{journal}{\bibinfo{title}{Viability of superoxide-containing
  radical pairs as magnetoreceptors}}.
\newblock {\emph{\JournalTitle{The Journal of chemical physics}}}
  \textbf{\bibinfo{volume}{151}}, \bibinfo{pages}{225101}
  (\bibinfo{year}{2019}).

\bibitem{Kattnig2017}
\bibinfo{author}{Kattnig, D.~R.}
\newblock \bibinfo{journal}{\bibinfo{title}{Radical-pair-based magnetoreception
  amplified by radical scavenging: resilience to spin relaxation}}.
\newblock {\emph{\JournalTitle{The Journal of Physical Chemistry B}}}
  \textbf{\bibinfo{volume}{121}}, \bibinfo{pages}{10215--10227}
  (\bibinfo{year}{2017}).

\bibitem{Kattnig2017b}
\bibinfo{author}{Kattnig, D.~R.} \& \bibinfo{author}{Hore, P.}
\newblock \bibinfo{journal}{\bibinfo{title}{The sensitivity of a radical pair
  compass magnetoreceptor can be significantly amplified by radical
  scavengers}}.
\newblock {\emph{\JournalTitle{Scientific reports}}}
  \textbf{\bibinfo{volume}{7}}, \bibinfo{pages}{1--12} (\bibinfo{year}{2017}).

\bibitem{Raz2005}
\bibinfo{author}{Raz, A.}
\newblock \bibinfo{journal}{\bibinfo{title}{Perspectives on the efficacy of
  antidepressants for child and adolescent depression}}.
\newblock {\emph{\JournalTitle{PLoS Med}}} \textbf{\bibinfo{volume}{3}},
  \bibinfo{pages}{e9} (\bibinfo{year}{2005}).

\bibitem{Forlenza2012}
\bibinfo{author}{Forlenza, O.~V.}, \bibinfo{author}{de~Paula, V.~J.},
  \bibinfo{author}{Machado-Vieira, R.}, \bibinfo{author}{Diniz, B.~S.} \&
  \bibinfo{author}{Gattaz, W.~F.}
\newblock \bibinfo{journal}{\bibinfo{title}{Does lithium prevent alzheimer’s
  disease?}}
\newblock {\emph{\JournalTitle{Drugs \& aging}}} \textbf{\bibinfo{volume}{29}},
  \bibinfo{pages}{335--342} (\bibinfo{year}{2012}).

\bibitem{Forlenza2014}
\bibinfo{author}{Forlenza, O.~V.}, \bibinfo{author}{De-Paula, V. d. J.~R.} \&
  \bibinfo{author}{Diniz, B.}
\newblock \bibinfo{journal}{\bibinfo{title}{Neuroprotective effects of lithium:
  implications for the treatment of alzheimer’s disease and related
  neurodegenerative disorders}}.
\newblock {\emph{\JournalTitle{ACS chemical neuroscience}}}
  \textbf{\bibinfo{volume}{5}}, \bibinfo{pages}{443--450}
  (\bibinfo{year}{2014}).

\bibitem{Mahadik1996}
\bibinfo{author}{Mahadik, S.~P.} \& \bibinfo{author}{Mukherjee, S.}
\newblock \bibinfo{journal}{\bibinfo{title}{Free radical pathology and
  antioxidant defense in schizophrenia: a review}}.
\newblock {\emph{\JournalTitle{Schizophrenia research}}}
  \textbf{\bibinfo{volume}{19}}, \bibinfo{pages}{1--17} (\bibinfo{year}{1996}).

\bibitem{Clay2010}
\bibinfo{author}{Clay, H.~B.}, \bibinfo{author}{Sillivan, S.} \&
  \bibinfo{author}{Konradi, C.}
\newblock \bibinfo{journal}{\bibinfo{title}{Mitochondrial dysfunction and
  pathology in bipolar disorder and schizophrenia}}.
\newblock {\emph{\JournalTitle{International Journal of Developmental
  Neuroscience}}} \textbf{\bibinfo{volume}{29}}, \bibinfo{pages}{311--324}
  (\bibinfo{year}{2011}).

\bibitem{Snyder2013}
\bibinfo{author}{Gao, W.-J.} \& \bibinfo{author}{Snyder, M.~A.}
\newblock \bibinfo{journal}{\bibinfo{title}{Nmda hypofunction as a convergence
  point for progression and symptoms of schizophrenia}}.
\newblock {\emph{\JournalTitle{Frontiers in cellular neuroscience}}}
  \textbf{\bibinfo{volume}{7}}, \bibinfo{pages}{31} (\bibinfo{year}{2013}).

\bibitem{Jenner2003}
\bibinfo{author}{Jenner, P.}
\newblock \bibinfo{journal}{\bibinfo{title}{Oxidative stress in parkinson's
  disease}}.
\newblock {\emph{\JournalTitle{Annals of Neurology: Official Journal of the
  American Neurological Association and the Child Neurology Society}}}
  \textbf{\bibinfo{volume}{53}}, \bibinfo{pages}{S26--S38}
  (\bibinfo{year}{2003}).

\bibitem{Kent2009}
\bibinfo{author}{Kent, S.~T.} \emph{et~al.}
\newblock \bibinfo{journal}{\bibinfo{title}{Effect of sunlight exposure on
  cognitive function among depressed and non-depressed participants: a regards
  cross-sectional study}}.
\newblock {\emph{\JournalTitle{Environmental Health}}}
  \textbf{\bibinfo{volume}{8}}, \bibinfo{pages}{1--14} (\bibinfo{year}{2009}).

\bibitem{LeGates2014}
\bibinfo{author}{LeGates, T.~A.}, \bibinfo{author}{Fernandez, D.~C.} \&
  \bibinfo{author}{Hattar, S.}
\newblock \bibinfo{journal}{\bibinfo{title}{Light as a central modulator of
  circadian rhythms, sleep and affect}}.
\newblock {\emph{\JournalTitle{Nature Reviews Neuroscience}}}
  \textbf{\bibinfo{volume}{15}}, \bibinfo{pages}{443--454}
  (\bibinfo{year}{2014}).

\bibitem{Srinivasan2006}
\bibinfo{author}{Srinivasan, V.} \emph{et~al.}
\newblock \bibinfo{journal}{\bibinfo{title}{Melatonin in mood disorders}}.
\newblock {\emph{\JournalTitle{The World Journal of Biological Psychiatry}}}
  \textbf{\bibinfo{volume}{7}}, \bibinfo{pages}{138--151}
  (\bibinfo{year}{2006}).

\bibitem{LMcCOLL2001}
\bibinfo{author}{McColl, S.~L.} \& \bibinfo{author}{Veitch, J.~A.}
\newblock \bibinfo{journal}{\bibinfo{title}{Full-spectrum fluorescent lighting:
  a review of its effects on physiology and health}}.
\newblock {\emph{\JournalTitle{Psychological medicine}}}
  \textbf{\bibinfo{volume}{31}}, \bibinfo{pages}{949--964}
  (\bibinfo{year}{2001}).

\bibitem{Pail2011}
\bibinfo{author}{Pail, G.} \emph{et~al.}
\newblock \bibinfo{journal}{\bibinfo{title}{Bright-light therapy in the
  treatment of mood disorders}}.
\newblock {\emph{\JournalTitle{Neuropsychobiology}}}
  \textbf{\bibinfo{volume}{64}}, \bibinfo{pages}{152--162}
  (\bibinfo{year}{2011}).

\bibitem{Winkler2006}
\bibinfo{author}{Winkler, D.}, \bibinfo{author}{Pjrek, E.},
  \bibinfo{author}{Iwaki, R.} \& \bibinfo{author}{Kasper, S.}
\newblock \bibinfo{journal}{\bibinfo{title}{Treatment of seasonal affective
  disorder}}.
\newblock {\emph{\JournalTitle{Expert review of neurotherapeutics}}}
  \textbf{\bibinfo{volume}{6}}, \bibinfo{pages}{1039--1048}
  (\bibinfo{year}{2006}).

\bibitem{Penckofer2010}
\bibinfo{author}{Penckofer, S.}, \bibinfo{author}{Kouba, J.},
  \bibinfo{author}{Byrn, M.} \& \bibinfo{author}{Estwing~Ferrans, C.}
\newblock \bibinfo{journal}{\bibinfo{title}{Vitamin d and depression: where is
  all the sunshine?}}
\newblock {\emph{\JournalTitle{Issues in mental health nursing}}}
  \textbf{\bibinfo{volume}{31}}, \bibinfo{pages}{385--393}
  (\bibinfo{year}{2010}).

\bibitem{Spedding2014}
\bibinfo{author}{Spedding, S.}
\newblock \bibinfo{journal}{\bibinfo{title}{Vitamin d and depression: a
  systematic review and meta-analysis comparing studies with and without
  biological flaws}}.
\newblock {\emph{\JournalTitle{Nutrients}}} \textbf{\bibinfo{volume}{6}},
  \bibinfo{pages}{1501--1518} (\bibinfo{year}{2014}).

\bibitem{Parker2017}
\bibinfo{author}{Parker, G.~B.}, \bibinfo{author}{Brotchie, H.} \&
  \bibinfo{author}{Graham, R.~K.}
\newblock \bibinfo{journal}{\bibinfo{title}{Vitamin d and depression}}.
\newblock {\emph{\JournalTitle{Journal of affective disorders}}}
  \textbf{\bibinfo{volume}{208}}, \bibinfo{pages}{56--61}
  (\bibinfo{year}{2017}).

\bibitem{Uberti2016}
\bibinfo{author}{Uberti, F.} \emph{et~al.}
\newblock \bibinfo{journal}{\bibinfo{title}{Protective effects of vitamin d 3
  on fimbrial cells exposed to catalytic iron damage}}.
\newblock {\emph{\JournalTitle{Journal of ovarian research}}}
  \textbf{\bibinfo{volume}{9}}, \bibinfo{pages}{1--10} (\bibinfo{year}{2016}).

\bibitem{Uberti2017}
\bibinfo{author}{Uberti, F.} \emph{et~al.}
\newblock \bibinfo{journal}{\bibinfo{title}{Biological effects of combined
  resveratrol and vitamin d3 on ovarian tissue}}.
\newblock {\emph{\JournalTitle{Journal of ovarian research}}}
  \textbf{\bibinfo{volume}{10}}, \bibinfo{pages}{1--14} (\bibinfo{year}{2017}).

\bibitem{Hameroff2014}
\bibinfo{author}{Hameroff, S.} \& \bibinfo{author}{Penrose, R.}
\newblock \bibinfo{journal}{\bibinfo{title}{Consciousness in the universe: A
  review of the ‘orch or’theory}}.
\newblock {\emph{\JournalTitle{Physics of life reviews}}}
  \textbf{\bibinfo{volume}{11}}, \bibinfo{pages}{39--78}
  (\bibinfo{year}{2014}).

\bibitem{Hameroff2014b}
\bibinfo{author}{Hameroff, S.~R.}, \bibinfo{author}{Craddock, T.~J.} \&
  \bibinfo{author}{Tuszynski, J.~A.}
\newblock \bibinfo{journal}{\bibinfo{title}{Quantum effects in the
  understanding of consciousness}}.
\newblock {\emph{\JournalTitle{Journal of integrative neuroscience}}}
  \textbf{\bibinfo{volume}{13}}, \bibinfo{pages}{229--252}
  (\bibinfo{year}{2014}).

\bibitem{Simon2019}
\bibinfo{author}{Simon, C.}
\newblock \bibinfo{journal}{\bibinfo{title}{Can quantum physics help solve the
  hard problem of consciousness?}}
\newblock {\emph{\JournalTitle{Journal of Consciousness Studies}}}
  \textbf{\bibinfo{volume}{26}}, \bibinfo{pages}{204--218}
  (\bibinfo{year}{2019}).

\bibitem{Adams2020}
\bibinfo{author}{Adams, B.} \& \bibinfo{author}{Petruccione, F.}
\newblock \bibinfo{journal}{\bibinfo{title}{Quantum effects in the brain: A
  review}}.
\newblock {\emph{\JournalTitle{AVS Quantum Science}}}
  \textbf{\bibinfo{volume}{2}}, \bibinfo{pages}{022901} (\bibinfo{year}{2020}).

\bibitem{Cifra2014}
\bibinfo{author}{Cifra, M.} \& \bibinfo{author}{Posp{\'\i}{\v{s}}il, P.}
\newblock \bibinfo{journal}{\bibinfo{title}{Ultra-weak photon emission from
  biological samples: definition, mechanisms, properties, detection and
  applications}}.
\newblock {\emph{\JournalTitle{Journal of Photochemistry and Photobiology B:
  Biology}}} \textbf{\bibinfo{volume}{139}}, \bibinfo{pages}{2--10}
  (\bibinfo{year}{2014}).

\bibitem{Kumar2016}
\bibinfo{author}{Kumar, S.}, \bibinfo{author}{Boone, K.},
  \bibinfo{author}{Tuszy{\'n}ski, J.}, \bibinfo{author}{Barclay, P.} \&
  \bibinfo{author}{Simon, C.}
\newblock \bibinfo{journal}{\bibinfo{title}{Possible existence of optical
  communication channels in the brain}}.
\newblock {\emph{\JournalTitle{Scientific reports}}}
  \textbf{\bibinfo{volume}{6}}, \bibinfo{pages}{36508} (\bibinfo{year}{2016}).

\bibitem{Berkovitch2020}
\bibinfo{author}{Berkovitch, L.} \emph{et~al.}
\newblock \bibinfo{journal}{\bibinfo{title}{Disruption of conscious access in
  psychosis is associated with altered structural brain connectivity}}.
\newblock {\emph{\JournalTitle{Journal of Neuroscience}}}
  \textbf{\bibinfo{volume}{41}}, \bibinfo{pages}{513--523}
  (\bibinfo{year}{2021}).

\bibitem{Neese2011}
\bibinfo{author}{Neese, F.}
\newblock \bibinfo{journal}{\bibinfo{title}{The orca program system}}.
\newblock {\emph{\JournalTitle{Wiley Interdisciplinary Reviews: Computational
  Molecular Science}}} \textbf{\bibinfo{volume}{2}}, \bibinfo{pages}{73--78}
  (\bibinfo{year}{2012}).

\bibitem{Goerigk2011}
\bibinfo{author}{Goerigk, L.} \& \bibinfo{author}{Grimme, S.}
\newblock \bibinfo{journal}{\bibinfo{title}{A thorough benchmark of density
  functional methods for general main group thermochemistry, kinetics, and
  noncovalent interactions}}.
\newblock {\emph{\JournalTitle{Physical Chemistry Chemical Physics}}}
  \textbf{\bibinfo{volume}{13}}, \bibinfo{pages}{6670--6688}
  (\bibinfo{year}{2011}).

\bibitem{vanLenthe1996}
\bibinfo{author}{Van~Lenthe, E.~v.}, \bibinfo{author}{Snijders, J.} \&
  \bibinfo{author}{Baerends, E.}
\newblock \bibinfo{journal}{\bibinfo{title}{The zero-order regular
  approximation for relativistic effects: The effect of spin--orbit coupling in
  closed shell molecules}}.
\newblock {\emph{\JournalTitle{The Journal of chemical physics}}}
  \textbf{\bibinfo{volume}{105}}, \bibinfo{pages}{6505--6516}
  (\bibinfo{year}{1996}).

\bibitem{Marenich2009}
\bibinfo{author}{Marenich, A.~V.}, \bibinfo{author}{Cramer, C.~J.} \&
  \bibinfo{author}{Truhlar, D.~G.}
\newblock \bibinfo{journal}{\bibinfo{title}{Universal solvation model based on
  solute electron density and on a continuum model of the solvent defined by
  the bulk dielectric constant and atomic surface tensions}}.
\newblock {\emph{\JournalTitle{The Journal of Physical Chemistry B}}}
  \textbf{\bibinfo{volume}{113}}, \bibinfo{pages}{6378--6396}
  (\bibinfo{year}{2009}).

\end{thebibliography}


\section*{Acknowledgements}
The authors would like to thank Dennis Salahub, Wilten Nicola, Rishabh, Mansoor Askari, and Jordan Smith for their input, comments, and insights.The authors would like to acknowledge Compute Canada for its computing resources. This work was supported by the Natural Sciences and Engineering Research Council of Canada.

\section*{Author contributions statement}
H.ZH. performed calculations and modelling with feedback from C.S.; H.ZH. and C.S. wrote the paper; C.S. conceived and supervised the project.

\section*{Competing Interests}
The authors declare no competing interests.

\end{document}